\documentclass[epj]{svjour}
\usepackage{graphics}
\begin{document}
\title{A new lattice measurement for potentials between static $SU(3)$ sources}
\author{Sedigheh Deldar\thanks{e-mail: sdeldar@ut.ac.ir} 
}                     
%
%
\institute{Department of Physics, University of Tehran, P.O. Box 14395/547, Tehran 1439955961, Iran}

\date{Received: date / Revised version: date}
%
\abstract{
In this article, a new calculation of static potentials between sources of different representations in $SU(3)$ gauge group is presented. The results of author's previous study \cite{Deld00} at the smallest lattice spacing $a_{s}\simeq0.11$~ fm are shown to have been affected by finite volume effects. Within statistical errors, the new results obtained here are still in agreement with both, Casimir scaling and flux tube counting. There is also no contradiction to the results obtained in Ref.~ \cite{Bali00} which however exclude flux counting.
} 
%
\maketitle
\section{Introduction}
\label{intro}
Studying the potential between static sources in QCD is still an interesting
subject. At large distances, linearity of the potential leads to the confinement which is one of
the most challenging issues in QCD. There are various confinement hypotheses. The Casimir scaling, sine scaling, and flux tube counting are among those hypotheses which try to describe the potential between static sources in different representations. The question of the Casimir scaling/sine scaling/flux tube counting is essential for our understanding of the confinement mechanism and to clarify the correspondence between quantum field theories at a large number of colors and string theories on certain compactified manifolds. 

At short distances, potentials between two quarks can be calculated using perturbation theory. For this regime, called asymptotic freedom, the coupling constant is small enough to use perturbation theory. As the distance between quarks increases, a tube of chromoelectric flux is formed between them and a potential proportional to the quark separation is expected. Perturbation theory dose not work for this region. Lattice gauge theory is one of the most successful method which works well for this low energy regime where the potential may be defined as:
\begin{equation}
V(r)\simeq A/r + Kr + C
\label{potential}
\end{equation}
where $r$ and $K$ indicate the quark separation and the string tension, respectively. The first term shows the Coulombic potential which is the result of one gluon exchange at short distances. The confinement is understood from the second term which shows that the potential between the pair of quark-antiquark increases by distance. Quite a few lattice studies have devoted to the calculations of string tensions in the closed string sector (torelons) \cite{Luci03}, yet only two authors, S. Deldar \cite{Deld00} and G. Bali \cite{Bali00}, have recently systematically studied the situation in the open string sector and have measured the potentials between static sources for a variety of representations. Both calculations have shown good agreement with the Casimir scaling, especially based on Bali's paper, within an accuracy of $5$ percent, no violation to the Casimir scaling is detected. Here, the Casimir scaling means that the string tension for each
representation is to be roughly proportional to the eigenvalue of the quadratic Casimir operator in that representation. The Casimir 
scaling regime is expected to exist for intermediate distances, perhaps extending from the onset of confinement to the onset of screening 
\cite{Fabe97}. There is another argument about the linear part of the potential at intermediate distances which claims that the string tension at this region is proportional to the number of fundamental flux tubes embedded
into the representation \cite{shif00}. The fundamental flux or string is the one that connects a fundamental heavy quark with an anti-quark. This idea is called ``flux tube counting''. In general, at very large $N$ strings do not interact , therefore, one would expect that the tension of meta-stable strings is proportional to the number of flux tubes. Of course, there are $\frac{1}{N}$ corrections. And then, if one waits for long enough, these meta-stable strings will decay into the stable strings with the given $N$-ality, whose tension is likely to be described by the sine formula or the Casimir scaling in some approximation which is not yet known so far. 

The confining string with $N$-ality $k$, is usually called k-string and $\sigma_{k}$ is the corresponding string tension. For $SU(N)$ gauge sources at large distances, where the strings are stable and do not decay, there exist some different theories about the ratio of $(\frac{\sigma_{k}}{\sigma_{f}})$ where $\sigma_{f}$ and $\sigma_{k}$ are the string tensions of the fundamental quarks and k-sources, respectively. The most trivial idea is, to assume that the total flux is carried by $k$ independent fundamental tubes:
\begin{equation}
\sigma_{k}=k\sigma_{f}
\label{largeN}
\end{equation}
Because of charge conjugation, $\sigma_{k}=\sigma_{N-k}$. Thus, for $SU(3)$ gauge group, $\sigma_{2}=\sigma_{1}$ and one universal string tension is obtained. In this case, string tensions of non-zero triality representations will be equal to the string tension of fundamental quarks at large distances.  Asymptotic Casimir scaling is another theory about k-string ratios \cite{Ambj84}:
\begin{equation}
\frac{\sigma_{k}}{\sigma_{f}}=\frac{k(N-k)}{N-1}
\label{Cas}
\end{equation}
 Another hypothesis is based on calculations in brane M.theory \cite{Doug95} or Sine-law scaling:
\begin{equation}
\frac{\sigma_{k}}{\sigma_{f}}=\frac{sin\frac{k\pi}{N}}{sin\frac{\pi}{N}}
\label{sin}
\end{equation}
In the large $N$ limit, the Casimir and sine-law scaling will be the same and equal to $k$.

Back to $SU(3)$ gauge group, at large distances, larger than $1.2$ fm, where the potential between two sources is large enough, a pair
of adjoint sources releases from the vacuum and string breaking may happen. Then, based on the triality
of the representation, we expect to see screening or change of the slope
of the potential to the slope of the potential of the fundamental representation. Since there is only one independent string tension for $SU(3)$ gauge field which is the string tension of the fundamental representation, potentials of zero triality representations like the adjoint representation are expected to be screened at large enough distances which means that the string tensions will be equal to zero and for non-zero triality representations, one expects to see the string tension of the fundamental representation.

Although in this study, potentials for distances larger than $1.2$ fm are calculated, no string breaking and therefore no sign of the screening or the change of the slope of the potentials to that of the fundamental representation is observed. As mentioned in the previous paper \cite{Deld00}, it is probably because of the fact that the Wilson loops do not couple well to the screened representations. As will be explained in the next section, potentials are calculated by measuring the Wilson loops. Therefore, through this article, the intermediate string tensions and two of the theories which may be applied to this region are discussed. These theories are flux tube counting and the Casimir scaling for intermediate distances. Equations \ref{largeN}, \ref{Cas} and \ref{sin} for stable strings which may be applied to large distances are not investigated. I should mention that no comparison with MQCD or sine scaling could be done, since
as far as the author knows, no calculations in that framework are done for the meta-stable strings \cite{Shif-tep}. There are comparison with both the Casimir scaling and sine-scaling in the studies which measure the string tensions in the closed string sector (torelons) \cite{Luci03}.

In spite of the good agreement between author's previous calculations and the Casimir scaling, the results of one of the lattices have not been scaled well with the others. The simulations were done
on a couple of anisotropic lattices with spatial
lattice spacings of $0.43$, $0.25$ and $0.11$ fm. The results for lattice spacing $0.43$ fm and $0.25$ fm
have been in good agreement but the data for the finest lattice have not been scaled well
with others, especially for higher representations. A new lattice spacing has
been examined to probe the disagreement. This measurement is really important
since in the author's previous calculations, the non-scaled potentials were the ones
belonged to the finer lattice.
In fact, in order to get the continuum, one has to use finer lattices; therefore, the results of finer lattice should be more reliable. This is in contrary to the author's previous calculations where the finer lattice has not behaved properly. In
this paper, it will be shown that the problem with the finer lattice has been the significant smaller volume compared with others, which
leads to the finite volume effect error that is one of the most important errors of lattice gauge theory calculations.  On the other hand, with the new lattice
spacing, $a_{s}=0.19$ fm, the lattice is still considered fine and the volume is
large enough to not encounter the finite volume error.

The hypotheses of the Casimir scaling and the flux tube counting for the meta-stable strings are also investigated and it will be shown that string tensions are roughly in agreement with both theories.
                                                                               
The paper is organized as the following: in
section two, the Wilson loops and the potentials and in section three, the action and the lattice
are discussed. Results, scaling behavior, the string tensions and their features, and the conclusion are discussed in the next sections, respectively.

\section{Wilson loops and the static potentials}
\label{sec:1}

The potential between two static quarks is found by measuring the Wilson loop
and looking for the area law fall-off for large t:
\begin{equation}
W(r,t)\simeq \exp[-V(r)t]
\label{Wilson}
\end{equation}
$W(r,t)$ is the Wilson loop as a function of $r$, the distance between two sources,
and $t$, the propagation time.
The potential at distance $r$ is determined from the asymptotic behavior
of the Wilson loop, $W(r,t)$:
\begin{equation}
V(r)\simeq \lim_{t\rightarrow{+\infty}} \ln(\frac{W(r,t)}{W(r,t+1)})
\label{lnWil}
\end{equation}
The string tension and the coefficient of the Coulombic term may be obtained by
fitting $V(r)$ to equation \ref{potential}. $V(r)$'s of different $r$ are obtained from equation \ref{lnWil}.  
Wilson loops of higher representations ($R$'s), $W_{R}$, are calculated from the fundamental 
Wilson loop, $U$, by the tensor product method. Trace of $W_{R}$ for
representations 6, 8, 10, 15 symmetric, 15 antisymmetric and 27 which are
the representations studied in this paper are as the following:
\begin{equation}
tr(W_{6}) = 1/2~ [~(trU)^2 + trU^2)~ ],
\label{r6}
\end{equation}
\begin{equation}
tr(W_{8}) = trU^\star trU -1,
\label{r8}
\end{equation}
\begin{eqnarray}
tr(W_{10}) = 1/6~[~(trU)^3 + 2(trU^3) + 3trUtrU^2~ ],
\label{r10}
\end{eqnarray}
\begin{eqnarray}
tr(W_{15s}) = 1/24~[~ (trU)^4 + 6(trU)^2trU^2  \nonumber \\
              +8trU(trU^3)
              +3(trU^2)^2 + 6trU^4 ~],
\label{r15s}
\end{eqnarray}
\begin{eqnarray}
tr(W_{15a}) = 1/2trU^\star~[~(trU)^2+ trU^2]-trU,
\label{r15a}
\end{eqnarray}
\begin{eqnarray}
tr(W_{27})=1/4[trU^2+(trU)^2]~[(tr(U^\star)^2)+(trU^\star)^2~]\nonumber \\
-trUtrU^\star,
\label{r27}
\end{eqnarray}

\section{Action and the lattice}
\label{sec:2}
                                                                                
A $16^3\times24$ lattice has been used for this new measurement. The coupling
constant is $2.7$ and the ratio of the spatial lattice spacing to the temporal
one, $\xi=\frac{a_{s}}{a_{t}}$, is equal to $2$. The improved action used for
this lattice is \cite{Pead97}:
\begin{eqnarray}
S = \beta{\{\frac{5}{3}\frac{\Omega_{sp}}{\xi u_{s}^4}+
\frac{4}{3} \frac{\xi\Omega_{tp}}{u_{s}^2u_{t}^2}-
\frac{1}{12}\frac{\Omega_{sr}}{\xi u_{s}^6}-
\frac{1}{12}\frac{\xi\Omega_{str}}{u_{s}^4u_{s}^2 }}\}.
\label{anisot}
\end{eqnarray}
where $\beta=6/g^2$, $g$ is the QCD coupling, and $\xi$ is the aspect ratio
($\xi=a_{s}/a_{t}$ at tree level in perturbation theory). $\Omega_{sp}$
and $\Omega_{tp}$ include the sum over spatial and temporal plaquettes;
$\Omega_{sr}$ and $\Omega_{str}$ include the sum over $2\times1$ spatial
rectangular and short temporal rectangular (one temporal and two spatial
links), respectively. For $a_{t} \ll a_{s}$ the discretization error of
this action is $O(a_{s}^4,a_{t}^2,a_{t}a_{s}^2)$. The coefficients are
determined at tadpole-improved tree level \cite{Lepa93}. The spatial
mean link, $u_{s}$ is given by:
\begin{equation}
\langle\frac{1}{3}ReTrP_{ss'}\rangle ^\frac{1}{4},
\end{equation}
where $P_{ss'}$ denotes the spatial plaquette. In general the temporal
link $u_{t}$, can be determined from:
\begin{equation}
u_{t} = \frac{\sqrt { \frac{1}{3} \langle Re Tr P_{st}\rangle}}{u_{s}},
\label{ut}
\end{equation}
where $P_{st}$ is the spatial-temporal plaquette. When
$a_{t} \ll a_{s}$, $u_{t}$, the temporal mean link can be fixed to
$u_{t}=1$, since its value in perturbation theory differs from unity by
$O(\frac{a_{t}^2}{a_{s}^2})$. To minimize the excited state contamination in correlation functions,
spatial links are smeared \cite{Alba87}, (APE smearing).

\vspace{30pt}

\section{Results}
\label{Results}

\begin{figure}
\vspace{50pt}
\resizebox{0.47\textwidth}{!}{
\includegraphics{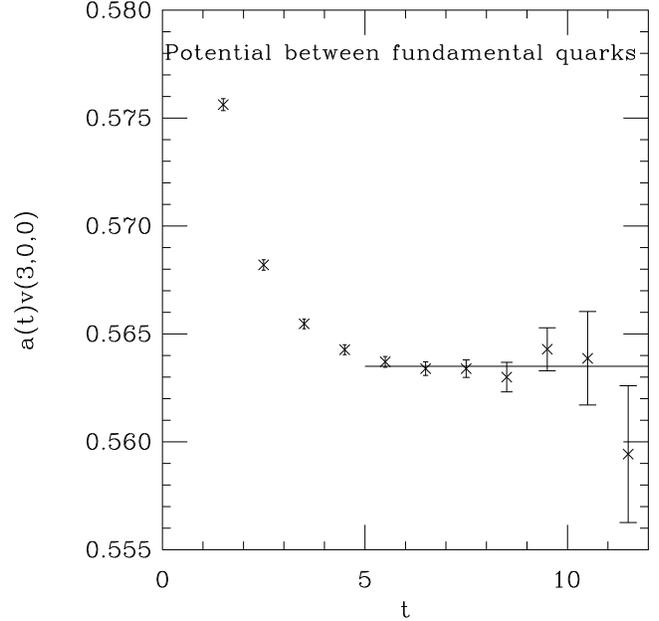}}
\caption{Potential versus 
$t$ for the fundamental representation. The fit range is shown by the solid line.}
\label{fig:fund}
\end{figure}

\begin{figure}
\vspace{50pt}
\resizebox{0.45\textwidth}{!}{
\includegraphics{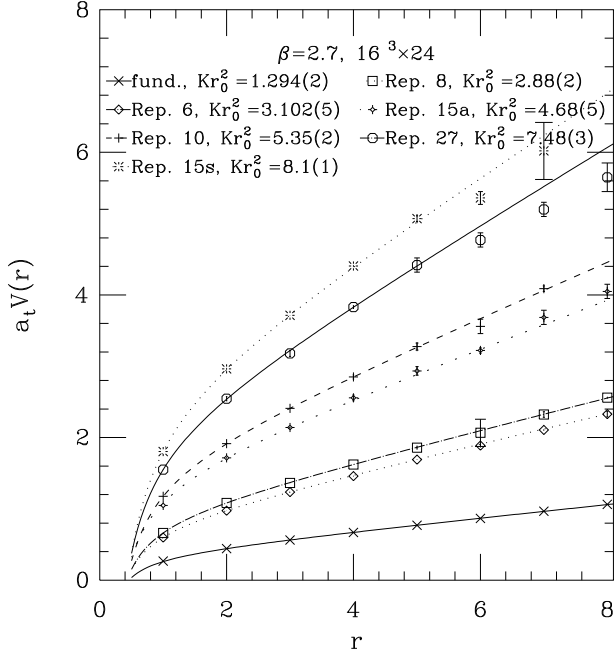}}
\caption{
Potentials for the fundamental, 6, 8, 10, 15a, 15s and 27
representations. The fits are based on 12800 measurements.
Rough agreement with the Casimir scaling is observed at the intermediate distances.
$Kr_{0}$, the dimensionless string tension for each representation is indicated
in the plot. $r_{0}$ is the hadronic scale which is defined in terms of the
force between static quarks at intermediate distance.
}
\label{fig:rn}
\end{figure}
                                                                                
Using equation \ref{lnWil} and measuring the Wilson loops for a variety of $t$'s for fixed $r$'s, the potential $V(r)$ can be found. This process is repeated for several $r$'s and the optimum $V(r)$ for each $r$ is extracted from plots like figure \ref{fig:fund}. More details may be found in \cite{Deld00}. Figure \ref{fig:rn} shows $V(r)$ versus $r$ for all representations using the
new coupling
constant. The potentials have been fitted to equation \ref{potential} which has
a linear
plus a Coulombic term. The error bars on the points are the sum in quadratic
of statistical and systematic errors. The systematic errors are due to the
change of the fit range of $V$ versus $r$. From the plot, one can see that the
potentials are linear at intermediate distances and thus quarks are confined
at this regime.
At short distances, potentials are proportional to
$\frac{1}{r}$. The slopes of the potentials of higher representations are expected to decrease
at large distances so that zero triality representations are screened and the
slope of the potentials of quarks with non-zero triality representations changes
to the slope of the potential between quarks in the fundamental representation.
Change of the slope for the potential between gluons (quarks in the adjoint representation) is
expected to happen at $r\simeq1.2$ fm, where the potential is equal to the potential
energy of gluelumps. The gluon anti-gluon released from the vacuum, couple to the initial sources and screening would happen. For the coupling constant used in this calculation
which is $\beta=2.7$, the
lattice spatial spacing is about $0.19$ fm and the maximum lattice
distance is $r=8$. Thus the maximum physical distance between static sources
is about $1.5$ fm which is not really large enough to observe significant bending
of screened potentials. In fact, even in the previous measurements where the maximum
lattice distance was about $2.2$ fm, this change of the slope of the potential or string breaking has not
been observed. As discussed in reference \cite{Deld00}, it is probably because
of the well known fact that the Wilson loops do not couple well to screened states. It should be mentioned that the errors of the large distances potentials are large enough to not have a significant role in the fitting.

\begin{figure}[]
\vspace{50pt}
\resizebox{0.45\textwidth}{!}{
\includegraphics{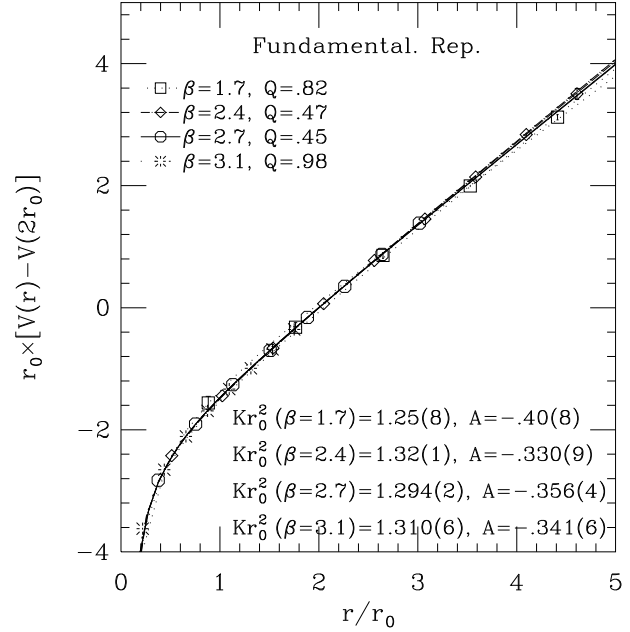}}
\caption{The static quark potential $V(r)$ in terms of hadronic scale
$r_{0}$ for the fundamental representation. Potentials of all four measurements
are in agreement. $Q$ represents the confidence level of the fit.}
\label{fig:Morn_f}
\end{figure}

\begin{figure}[]
\vspace{50pt}
\resizebox{0.45\textwidth}{!}{
\includegraphics{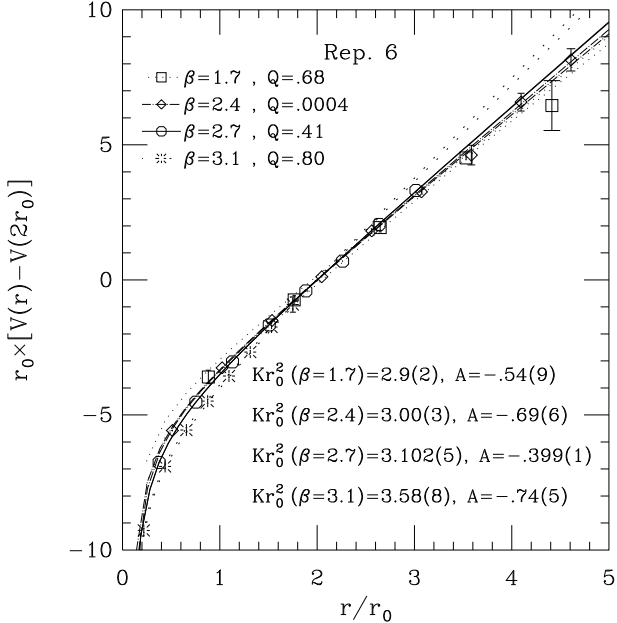}}
\caption{
Same as  figure \ref{fig:Morn_f} but for representation 6. Results from three lattices including the
new one, scale well. Potentials of $\beta=3.1$ do not scale since the
lattice spatial volume is significantly smaller than others and finite size error affects that measurement.
}
\label{fig:morn6}
\end{figure}
                                                                                
\section{Scaling behavior}
                                                                                
\begin{figure}[]
\vspace{50pt}
\resizebox{0.45\textwidth}{!}{
\includegraphics{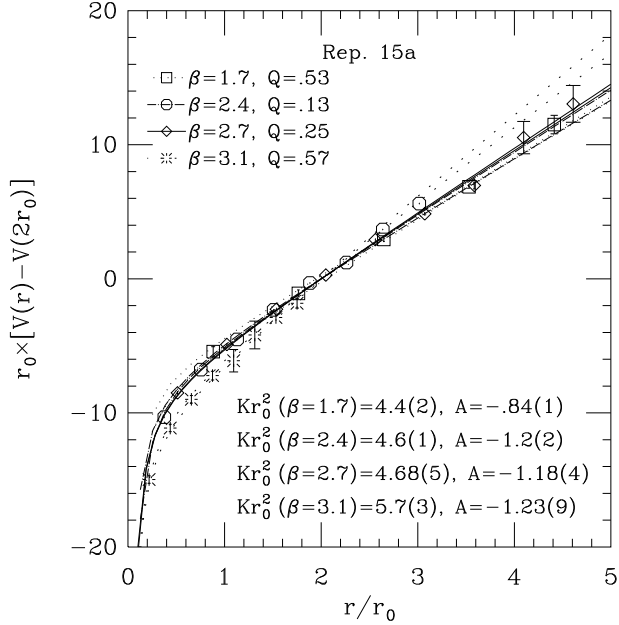}}
\caption{Same as figure \ref{fig:morn6} but for representation 15a.}   
\label{fig:morn15}
\end{figure}

The actual measurement by lattice simulations is represented in lattice units. To convert the units into the physical units, one has to look for a
physical quantity with a known value. Then this reference quantity is measured
on the lattice and by comparing the two values, the lattice spacing is 
extracted in physical units. The hadronic scale $r_{0}$, defined in
terms of the force between static quarks at intermediate distance,
\cite{Somm94} can be used as the reference quantity:
\begin{equation}
[r^2dV/dr]_{r=r_{0}}=1.65
\label{scale}
\end{equation}
where $V(r)$ is the static quark potential in the fundamental representation.
The definition of Eq.\ (\ref{scale}) gives
$r_{0}\simeq 0.5$ fm in a phenomenological potential model.
$r_{0}^{-1}=410\pm 20$ MeV determined by C. Morningstar {\it et al.}
\cite{Morn97} is used in this study. To set the scale and find $a_{s}$, I use equation \ref{potential} and
the hadronic scale equation, equation \ref{scale}:
\begin{equation}
\frac{r_{0}}{a_{s}} = \sqrt{ \frac{1.65-A}{Ka_{s}^2} }.
\end{equation}
For the anisotropic lattice, $Ka_{s}a_{t}$ may be found from the fits. The input aspect ratio $\xi=\frac{a_{s}}{a_{t}}=2$ has been chosen since the difference
between the input value and lattice simulations is expected to vanish
in the continuum limit.
As shown by previous calculations, the good scaling behavior of the 
fundamental string tension is a good evidence that this assumption is correct.
It is now possible to show the results of different lattice simulations, using the scaled
potentials and lattice distances in terms of the hadronic scale $r_{0}$.
Figure \ref{fig:Morn_f} shows the static quark potential versus hadronic scale $r_{0}$ 
for the following four lattices: $10^3\times24$,
$18^3\times24$, $16^3\times24$, and $16^3\times24$ with spatial lattice
spacings $0.45$ fm, $0.25$ fm, $0.11$ fm and $0.19$ fm, respectively. The new
lattice is the one with lattice spacing equal to $0.19$ fm.
Proper scaling for all couplings for the fundamental representation is observed.
In contrast to the
fundamental representation where a good scaling is obtained for all four lattices, as
the dimension of representation increases, the finest lattice, $a_{s}=0.11$ fm, 
violates the scaling.
Figure ~\ref{fig:morn6} shows the potential for representation 6. The finest lattice does
not scale well but the data for other three measurements are in good agreement.
Figure ~\ref{fig:morn15}  
which shows the potentials for representations 15a, 
is another confirmation of this fact. The potentials of other 
representations, 8, 10, 15s and 27 show the same behavior.
In general, one expects to get the continuum by making the lattice spacing
finer. However, as figures ~\ref{fig:morn6} and \ref{fig:morn15}
indicates, the potentials of the finest lattice
do not show good scaling behavior. Probably this happens because of the
smaller lattice volume of this coupling. Table \ref{tab:lat-vol} shows lattice parameters
including lattice volumes and lattice spacings
for the four lattices of this study. The volume of the finest lattice,
$a_{s}=0.11$ fm, is significantly smaller than others. Therefore if one
does not want to encounter the error of finite volume effect should use
larger lattice volume.
This may happen by increasing the number of lattice points in each dimension
or increasing the lattice spacing. The former increases the running time
and is not economic. Thus the second option has been chosen and $\beta=2.7$ is
selected which gives the larger lattice spacing, $0.19$ fm with compared to the
previous finer lattice where the lattice spacing was equal to $0.11$ fm.
It is still finer than the two others with lattice spacings $0.45$ fm and $0.25$ fm.
I recall that these are spatial lattice spacings and temporal lattice spacing has
been kept the same by choosing appropriate aspect ratios. Potentials from this new measurement scales with the two previous measurements.
                                                                                
Using this new coupling constant and the two previous ones: $\beta=1.7$ and
$\beta=2.4$, potentials between static sources are studied and parameters
of potentials are obtained. Table \ref{best-kr0} shows string tensions from
the three scaled measurements.
$Kr_{0}^2$, the dimensionless string tension of each lattice measurement and 
the best estimate for each representation are indicated.
The best estimate is found by the weighted average of the three lattice
measurements. The first error in the best string tension is the statistical
error (from the weighted average) and the second one is the systematic error
of discretization determined by the standard deviation. The ratio of the string
tension is proportional to the ratio of the Casimir scaling of the 7th column.
Table \ref{best-A} shows the coefficient of the Coulombic term obtained from each lattice
measurement. Again the ratio of the coefficient of each representation to that
of the fundamental representation is brought for comparison with the ratio of
the Casimir scaling. 

\begin{table*}[]
\setlength{\tabcolsep}{1.pc}
\caption{Lattice parameters for the four lattice measurements. $\beta$ is the
coupling constant; $a_{s}$ indicates the spatial lattice spacing and
$\xi$ shows the ratio of spatial spacing to the 
temporal one. 
Number of configurations is brought in the last column. The spatial lattice 
volume for the finer lattice is significantly smaller than others. Since
possibly, the finite volume effect error affects this measurement, it has been
excluded from further calculations.}
\label{tab:lat-vol}
\begin{center}
\begin{tabular}{lccccc}
\hline
Lattice & $\beta$  & $\xi=\frac{a_{s}}{a_{t}}$   & $a_{s} (fm)$ &  Spatial Volume ($fm^3$)   & No configurations
\\
\hline
$10^3\times24$ & $1.7$  & $5.0$   &  $.43$  & $79.5$ & $22400$
       \\
$18^3\times24$ &  $2.4$ & $3.0$   &  $.25$  & $91.1$ &
 $21620$    \\
$16^3\times24$  & $2.7$  & $2.0$   &  $.19$  & $28.1$ &
 $12800$    \\
$16^3\times24$ &  $3.1$  & $1.5$  &  $.11$  & $5.5$ &
 $18200$    \\
\hline
\end{tabular}
\end{center}
\end{table*}

\begin{table*}[]
\setlength{\tabcolsep}{.9pc}
\caption{String tensions in terms of $r_{0}$ for different coupling
constants, lattice sizes, and the best estimate. Ratios of string tension of
each representation to that of the fundamental are roughly qualitatively in agreement
with the ratios of the corresponding the Casimir numbers as well as the number of fluxes.
}
\label{best-kr0}
\begin{center}
\begin{tabular}{lccccccc}
\hline
Rep.  & $Kr_{0}^2(\beta=1.7)$   & $Kr_{0}^2(\beta=2.4)$ &
$Kr_{0}^2(\beta=2.7)$ &  best estimate & $\frac{k_{r}}{k_{f}}$ & $\frac{C_{r}}{C_{f}}$ &flux No.   \\
\hline
3   & $1.25(8)$ & $1.32(1)$ &  $1.294(2)$  & $1.295(2)(36)$ & $1$ &  $1$ &$1$
      \\
8   & $2.60(1)$ & $2.60(3)$  &  $2.88(2)$  & $2.65(1)(17)$ & $2.05(1)(14)$ & $2.25$ & $2$
      \\
6   & $2.9(2)$  & $3.00(3)$ &  $3.102(5)$  & $3.10(1)(16)$ & $2.39(1)(14)$ & $2.5$ &$2$
   \\
15a & $4.4(2)$  & $4.6(1)$ &  $4.68(5)$  & $4.65(4)(18)$ & $3.59(3)(17)$ & $4.0$ & $3$
   \\
10  & $4.9(3)$  & $5.4(2)$ &  $5.35(2)$  & $5.35(2)(32)$ & $4.13(2)(27)$ & $4.5$
& $3$
  \\
27  & $5.9(5)$  & $6.62(6)$  &  $7.48(3)$  & $7.3(1)(10)$ & $5.64(2)(79)$ & $6$ & $4$
     \\
15s & $7.1(5)$  & $7.6(2)$  &  $8.1(1)$  & $7.97(9)(67)$ & $6.15(7)(54)$ & $7$ & $4$
    \\
\hline
\end{tabular}
\end{center}
\end{table*}
                                                                                
\begin{table*}[]
\setlength{\tabcolsep}{1.pc}
\caption{Coulombic coefficients found by different lattice calculations
and the best estimate. Rough agreement with the Casimir ratios are observed.}
\label{best-A}
\begin{center}
\begin{tabular}{lcccccc}
\hline
Rep.  & $A(\beta=1.7)$   & $A(\beta=2.4)$ &
$A(\beta=2.7)$ &  best estimate & $\frac{A_{r}}{A_{f}}$ & $\frac{C_{r}}{C_{f}}$
  \\
  & $10^3\times24$  & $18^3\times24$ &  $16^3\times24$  & & \\
\hline
3   & $-.40(8)$ &  $-.330(9)$ &  $-.356(4)$  & $-.352(4)(37)$
     & $1$ & $1$   \\
8   & $-.60(5)$ &  $-.93(3)$  &  $-.69(1)$  & $-.71(1)(18)$
    & $2.02(3)(54)$  & $2.25$   \\
6   & $-.54(9)$  &  $-.69(6)$ &  $-.798(2)$  & $-.80(1)(20)$
    & $2.27(3)(61)$ & $2.5$ \\
15a & $-.84(1)$  &  $-1.2(2)$ &  $-1.18(4)$  & $-.86(2)(33)$
    & $2.44(6)(97)$ & $4.0$ \\
10  & $-.50(2)$  & $-.5(2)$ &  $-1.43(1)$  & $-1.24(1)(76)$
    & $3.5(1)(22)$ & $4.5$ \\
27  & $-1.9(5)$  &  $-1.71(6)$  &  $-1.82(1)$  & $-1.82(1)(96)$
    & $5.1(1)(28)$ & $6$  \\
15s & $-1.6(4)$  &  $-2.1(2)$  &  $-2.28(4)$  & $-2.27(4)(49)$
    & $6.5(1)(16)$ & $7$   \\
\hline
\end{tabular}
\end{center}
\end{table*}
                                                                                
\section{Discussion on string tensions}
\label{strings}
\begin{figure}[]
\vspace{50pt}
\resizebox{0.45\textwidth}{!}{
\includegraphics{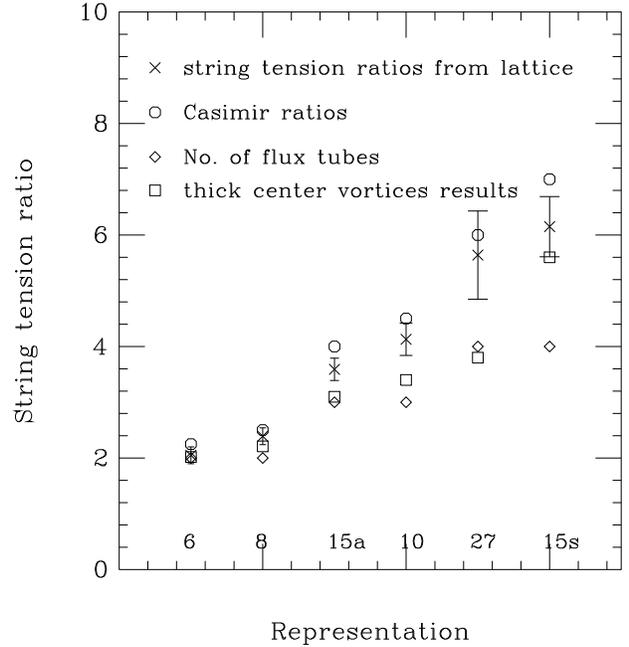}}
\caption{Ratios of string tensions of SU(3) quarks of
different representations to the string tension of quarks in the fundamental
representation are plotted. Considering the lattice data errors, a rough agreement with both
the Casimir scaling and flux tube counting is observed. String ratios obtained from
thick center vortices are also brought for comparison. }
\label{fig:latfig}
\end{figure}

Lattice calculations \cite{Deld00} and \cite{Bali00}   show that the potentials
between static $SU(3)$ sources are linear and proportional to the Casimir
operator of each representation, means  proportional to the eigenvalue of the quadratic Casimir operator of that representation. The proportionality of the potentials
with the Casimir operator is expected at short distances where the force between
heavy quarks may be described by one gluon exchange. But this behavior has
not been understood for intermediate distances even though it is observed not
only for $SU(3)$ but also for all $SU(N)$ gauge groups (examples are references \cite{Luci03} and \cite{Shif03} and references in them). Another
scenario which tries to explain the linear potentials at intermediate distances,
is flux tube counting. The idea is that the string tension of higher 
representation sources can be obtained by multiplying the number of fundamental
strings times the string tension of the quarks in the fundamental representation
\cite{shif00}. A fundamental string is a string which connects a fundamental 
heavy quark to an antiquark. The last column of table \ref{best-kr0} shows the number of the fundamental fluxes of each representation.  
Figure ~\ref{fig:latfig} shows the data of lattice 
calculation and the thick center vortices model \cite{Deld05}. Thick center vorices model is one of the phenomenological models which tries to describe the behavior of the  linear part of the static sources potentials. Cross signs show the
string ratios obtained from lattice calculation of this paper. The Casimir ratios
and the number of fundamental flux tubes are indicated by circles and diamonds,
respectively. Square signs represent the string tensions ratios obtained from 
thick center vortices model. As it is observed,
lattice calculations and thick center vortices results agree qualitatively with both the Casimir
scaling and flux tube counting. The possible reasoning that the string tension is larger
than the number of fluxes is discussed in the second reference of \cite{shif00} and also in \cite{Deld05}. This might happen because at intermediate distances, the fundamental fluxes overlap and a positive energy is added to the binding energy of fluxes and makes the string
tension larger than the number of fluxes times the fundamental string tension. 

Since the error due to the 
hadronic scale uncertainty is not considered in our lattice data, the lattice
errors are larger than what is reported in the figure. In addition, the
potential is supposed to be measured from the area fall-off at large $t$ from equation
$W(r,t)\simeq \exp^{-V(r)t}$. Since for large $r$ values --- especially 
for higher representations --- the Wilson loops get too small for large $t$, 
and the error due to statistical fluctuations makes the measurements
meaningless; therefore, calculation of the potentials using smaller $t$'s is essential.
Even though a systematic error by changing the fit
range or by comparing with $V$ of smaller $r$'s is obtained for potential
of each representation, it seems that the string tension is overestimated expecting a larger error
 especially for higher representations.
Figure ~\ref{fig:fund} shows potential between fundamental
quarks versus $t$ for $r=3$, where $r$ indicates the lattice distance. Fitting range
is shown by the solid line. As mentioned before, for higher representations, 
especially for large $r$, large $t$'s could not be used. Looking at this plot,
it seems that one overestimates the potentials if one measures the potentials
using small $t$'s.  

\section{Conclusion}
\label{conc}

Using a new coupling constant, $SU(3)$ potentials between static sources for
a variety of representations are obtained. String tensions are found using
this new measurement and the author's previous calculations which have shown good
scaling behavior. The data for the finest lattice have been excluded since
by comparison, it is observed that finite volume effect destroys the measurement.

String tensions still remain qualitatively in agreement with both the Casimir scaling and flux tube counting. In fact, the errors of the author's lattice data of this study are still too large to discriminate between the two hypotheses.
The results of this study is in agreement with the  data of \cite{Bali00} which have found that the potentials are proportional to the Casimir scaling but, however, clearly exclude flux tube counting. Furthermore, since the Wilson loops do not couple well with screened representation, investigation about k-string picture for large distances or stable strings could not be done.

The author would like to emphasize that even though there are some evidences of 
proportionality of the string tensions with the Casimir scaling, at intermediate distances, for $SU(N)$
gauge groups, the Casimir scaling is still a puzzle. Understanding the physics of string tensions and confinement is still an open and interesting subject.

\section{Acknowledgment}
\label{Ack}
                                                                                
I am grateful to MILC collaboration for using their codes. I would like to thank
C. Bernard for his help in this work and M. Teper for useful discussions about MQCD theory. I would like to thank M. Shifman for all his helps, especially, his patience in answering my questions.
I am grateful to the research council of University of Tehran for
partial support of this study.
                                                                                


\begin{thebibliography}{99}
                                                                                
\bibitem{Deld00} S. Deldar, Phys. Rev. D62, p. 034509, 2000.
\bibitem{Bali00} G. Bali, Phys. Rev. D62, p. 114503, 2000.
\bibitem{Luci03} B. Lucini, M. Teper, Phys. Lett. B501, p. 128, 2001,
                 L. Del Debbio, H. H. Panagopoulos, P. Rossi, E. Vicari, 
                 Phys. Rev. D65, p. 021501, 2002,
                 L. Del Debbio, H. H. Panagopoulos, P. Rossi, E. Vicari,
                 JHEP 0201, p. 9, 2002,
                 D. Antonov, L. Del Debbio, JHEP 0312, p. 60, 2003,
                 B. Lucini, M. Teper, U. Wenger, JHEP 0406, p. 12, 2004.
\bibitem{Fabe97} M. Faber, J. Greensite, S. Olejn\'{i}k, Phys. Rev. D57 P. 2603 (1998).
\bibitem{shif00} C. Michael, hep-ph/9809211,
                 G. S. Bali, Phys. Rept. 343, p. 1, 2001,
                 A. Armoni, M. Shifman, Nucl. Phys. B671, p. 67, 2003.
\bibitem{Ambj84} J. Amjorn, P. Olesen, C. Peterson, Nucl. Phys. B240, p. 189, p. 533, 1984; B244, p. 262, 1984; Phys. Lett. B142, p. 410, 1984.
\bibitem{Doug95} M. Douglas and S. Shenker, Nucl. Phys. B447, p. 271, 1995,
                 A. Hanany, M. Strassler and A. Zaffaroni, Nucl. Phys. B513, p. 87, 1998.
\bibitem{Shif-tep} Many thanks to M. Shifman and M. Teper for answering my questions about MQCD or sine-scaling theory.
\bibitem{Pead97} C. Morningstar, M. Peardon, Phys. Rev. D56, p. 4043, 1997.
\bibitem{Lepa93} G.P. Lepage and P.B. Mackenzie, Phys. Rev. D48, p. 2250, 1993.
\bibitem{Alba87} M. Albanese {\it et al.}, Phys. Lett. B192, p. 163, 1987.
\bibitem{Somm94} R. Sommer, Nucl. Phys. B411, p. 839, 1994.
\bibitem{Morn97} C. Morningstar, Nucl. Phys. B (Proc. Suppl.) 53, p. 917, 1997.
\bibitem{Shif03} A. Armoni, M. Shifman, Nucl. Phys. B664, p. 233, 2003, Acta  Phys. Polon. B36, p. 3805, 2005.

\bibitem{Deld05} S. Deldar, JHEP 0101, p. 013, 2001, S. Deldar, S. Rafibakhsh, Eur. Phys. J. C42, p. 319, 2005.
                                                                                
\end{thebibliography}
\end{document}